# Rapid evolution of the most luminous galaxies during the first 900 million years

Rychard J. Bouwens & Garth D. Illingworth

*UCO/Lick Observatory and Department of Astronomy, University of California Santa Cruz, Santa Cruz, California 95064, USA.*

**The first 900 million years (Myr) to redshift $z\approx6$ (the first seven per cent of the age of the Universe) remains largely unexplored for the formation of galaxies. Large samples of galaxies have been found at $z\approx6$ (refs 1–4) but detections at earlier times are uncertain and unreliable. It is not at all clear how galaxies built up from the first stars when the Universe was ~300 Myr old ($z\approx12$–15) to $z\approx6$, just 600 Myr later. Here we report the results of a search for galaxies at $z\approx7$–8, about 700 Myr after the Big Bang, using the deepest near-infrared and optical images ever taken. Under conservative selection criteria we find only one candidate galaxy at $z\approx7$–8, where ten would be expected if there were no evolution in the galaxy population between $z\approx7$-8 and $z\approx6$. Using less conservative criteria, there are four candidates, where 17 would be expected with no evolution. This demonstrates that very luminous galaxies are quite rare 700 Myr after the Big Bang. The simplest explanation is that the Universe is just too young to have built up many luminous galaxies at $z\approx7$–8 by the hierarchical merging of small galaxies.**

High-redshift galaxies can be identified using the well-established "dropout" technique[5]. Absorption shortward of 91.2 nm and 121.6 nm by intervening neutral hydrogen results in a sharp change (a break) in the spectral energy distribution in redshifted galaxies and can be used to determine their redshift[6]. In the three years since the first galaxies were discovered in the reionization epoch[7,8] using this dropout technique, 500 galaxies[1] are now known at $z\approx6$ and the frontier for exploration has shifted to galaxies at $z\approx7$–8 and higher redshifts. Finding galaxies at redshifts beyond $z\approx6$ is very challenging, however, because the sources are extremely faint, and the high sky backgrounds make ground-based searches in the near-infrared part of the spectrum very difficult. Although near-infrared imaging from space largely overcomes this problem, we are limited by the areal coverage of the available instrumentation. For example, the Hubble Space Telescope (HST) Near Infrared Camera and Multiobject Spectrograph (NICMOS) instrument[9]—the only near-infrared imaging instrument in space—has a very small field of view (0.8 arcmin$^2$), which is just 7% of the area of the optical camera on HST, the Advanced Camera for Surveys (ACS). This has made it extremely time-consuming to survey large areas of sky for high-redshift sources.

Nonetheless, we were able to use the deep near-infrared NICMOS data over the Hubble Ultra Deep Field (HUDF)[10] to conduct the first significant search for $z\approx7$-8 galaxies[3,11]. The few candidates found were consistent with a slight increase in the volume density of luminous sources from $z\approx7$-8 to $z\approx6$ (from 700 Myr to 900 Myr after the Big Bang). However, owing to possible instrumental problems for some of the candidates and the large uncertainty caused by large-scale structure effects, we were not able to draw very strong conclusions about the evolution of the luminosity function, the ultraviolet (UV) luminosity density, and the star-formation rate from $z\approx$7-8 to $z\approx$6. Attempts to establish these quantities based on other data sets have also been quite uncertain[12].

Fortunately, sufficient NICMOS and ACS data are now available that we can firmly establish the density of luminous galaxies at $z\approx$7-8 and the evolution with respect to $z\approx$6. These data come from many years of primary and parallel observations with HST and include >~20 arcmin$^2$ of deep NICMOS data (reaching J and H band limits of ~27–28 AB mag at 5$\sigma$) that overlap with HST ACS data of comparable depth. A substantial fraction of these data are available over the two Great Observatories Origins Deep Survey fields[13] and permit a much more robust search for $z\approx$7-8 galaxies than was possible with the HUDF data alone—significantly increasing our search area and allowing us to average over the effects of large-scale structure. The search fields used for our $z\approx$7-8 selection are given in Supplementary Fig. 1 and Supplementary Table 1[10,13–15]. These extensive ACS and NICMOS data sets were processed using recently developed reduction software[16] (R.J.B. *et al.*, manuscript in preparation). After registration of these data onto the same frame, object catalogues were generated with SExtractor[17] from all 4$\sigma$ detections found in a $\chi^2$ image (generated from the J and H images). These catalogues were cleaned of spurious or other marginal detections by requiring objects to be 4.5$\sigma$ in the H band (0.3" diameter aperture) and the photometry performed, after smoothing the images to a consistent point spread function.

We then considered two different z-dropout selections on these catalogues. z-dropout selections identify star-forming galaxies at $z\approx$7–8 that show a strong flux break between the z and J bands due to the 121.6 nm absorption feature, but have very blue infrared colours (see Supplementary Fig. 2). Our first selection was a relatively conservative one that was designed to isolate a robust set of z-dropouts. Our second selection was a little more aggressive and designed to find a larger sample of candidates (albeit less secure) and to test the robustness of the results derived from our first selection. Applying these criteria to our photometric catalogues, only one z-dropout candidate made our conservative selection and four candidate z-dropouts made our less-conservative selection. There were two other sources in our search fields that satisfied our selection criteria, but they are very likely to be T-dwarf stars. All four $z\approx$7-8 candidates are shown in Fig. 1 and their properties are listed in Supplementary Table 2. A detailed discussion of our current selection is provided in the Supplementary Information.

What do these z-dropout candidates imply for the evolution of the rest-frame UV luminosity function? (The UV luminosity function describes the number density of galaxies versus UV luminosity and is usually parameterized as $\phi^* e^{-L/L^*}(L/L^*)^\alpha$ where $\phi^*$ is the normalization, $L^*$ is the characteristic luminosity, and $\alpha$ is the faint-end slope.) We compared the current numbers with what we might have expected if galaxies at $z\approx$7-8 were similar to those at $z\approx$6, just 200 million years later. We can compute the expected number by using our cloning software to artificially redshift the $z\approx$6 population over the redshift range of our sample, add them to our data, and then repeat the selection[18,19]. We estimated that we would find 10.1±3.5 z-dropouts using our conservative selection and 17.0±4.9 z-dropouts using our less conservative selection. This is clearly substantially more than the 1 and 4 z-dropout candidates found in our conservative and less-conservative selections, respectively, which suggests that there has been substantial evolution for galaxies that lie at the bright end of the luminosity function. For simple poissonian statistics, the present samples are inconsistent with no evolution in the galaxy population between $z\approx$7-8 and $z\approx$6, at 99.90% and 99.93% confidence, respectively.

To make a proper assessment of the significance of this result, however, we need to account for the effects of galaxy clustering ('large-scale structure'). Modelling the selection volume in our individual fields with a window of depth $\Delta z=1$ at $z\approx$7–8 and bias of 7 (see Supplementary Information for details), we estimate the field-to-field variance to be ~41% and ~49% for our ~5-arcmin$^2$ fields and 0.8-arcmin$^2$ fields, respectively[20,21]. Including the effect of large-scale structure increases the uncertainties on our expected numbers, that is, $10.1^{+3.1}_{-5.0}$ sources with our conservative search and $17.0^{+4.7}_{-7.2}$ sources for our less-conservative search. Again, this is inconsistent with no galaxy evolution, at 99.5% and 99.4% confidence, respectively, and indicates a dramatic drop in the overall prevalence of luminous galaxies over this short time interval. We estimate that the number density of luminous galaxies at $z\approx$7.4 is just $0.10^{+0.19}_{-0.07}$ times that at $z\approx$6 for our conservative search and $0.24^{+0.20}_{-0.12}$ times the $z\approx$6 value for our less-conservative selection. The two determinations here are consistent within the errors and suggest that our basic result is robust. For simplicity, we will use statistics from our more rigorous, conservative search for the remainder of this paper.

What does this deficit of luminous star-forming objects at $z\approx$7-8 tell us about evolution of the UV luminosity function? This deficit could be due to a change in the overall number density of galaxies or due to a change in luminosity of the brightest objects (or some combination thereof). Perhaps the simplest and most natural explanation, though, is through an evolution of the characteristic luminosity $L^*$. It is already well established that $L^*$ brightens significantly over the interval $z\approx$6 to $z\approx$3 (Fig. 2)[1,2], so it seems reasonable to imagine that this trend continues from $z\approx$7-8 to $z\approx$6. A relevant question to ask is how much fainter the characteristic luminosity at $z\approx$7-8 would need to be to reproduce the observed decline. There are two obvious ways to model this change: (1) vary $L^*$ while keeping the normalization $\phi^*=0.002$ Mpc$^{-3}$ and faint-end slope $\alpha=-1.73$ fixed at the $z\approx$6 values and (2) vary $L^*$ while keeping the product of $\phi^*$ and $L^*$ fixed (and $\alpha$ fixed) to maintain a constant luminosity density at $z\geq$6. The latter model is motivated by a number of recent results at high redshift, which suggest that the rest-frame UV luminosity density—and the total star-formation rate density—shows very mild evolution[1,22]. These latter observations include the high stellar masses found in many $z\approx$5-6 systems and large optical depths measured by the Wilkinson Microwave Anisotropy Probe ($\tau=0.09\pm0.03$), both of which suggest substantial high-redshift ($z>$6) star formation[23–28]. For the above models, we can evaluate the evolution in the characteristic luminosity needed to produce the factor-of-ten deficit in luminous galaxies at $z\approx$7-8 by repeating the Monte Carlo simulations described above. The end result is that we find that the characteristic luminosity at $z\approx$7-8 in the rest-frame UV (~1,900 Å) must be 1.1±0.4 mag fainter than $z{\approx}6$ to reproduce the observed result for our first set of assumptions and 1.5±0.4 mag fainter for our second (see Fig. 2). In both cases, the evolution is fairly dramatic and suggests that most galaxies at $z{\approx}7$-8 are much fainter than our survey limits.

The evolution observed here can best be presented in the terms of the rest-frame continuum UV luminosity density integrated down to our approximate magnitude limit $M_{AB}{\approx}{-}19.7$ mag and compared with similar determinations at other redshifts (Fig. 3)[1,22,29,30]. As is evident from Fig. 3, the luminosity density, and especially the star-formation rate (the extinction-corrected luminosity density), increases rapidly at early times and indicates that there is a rapid build-up in the most-luminous systems that continues from $z{\approx}7$-8 (700 Myr) to $z{\approx}4$ (1,600 Myr).

The simplest explanation for this large deficit of the most luminous galaxies at $z{\approx}7$–8 relative to that at $z{\approx}6$, just 200 Myr later, is that the Universe is simply too young to have built up many luminous systems, and that the majority of galaxies at $z{\approx}7$-8 have luminosities much fainter than $M_{AB}{\approx}{-}19.7$ mag. Hierarchical build-up would suggest such a result, but detailed modelling is needed to establish whether the rate of change measured here is consistent. Determining the nature of the $z{\approx}7$–8 star-forming population will be a challenge, requiring significantly deeper observations. The HST Wide-Field Camera 3 will provide a near-term opportunity, but ultimate elucidation of the galaxy buildup in the first 500 Myr awaits the James Webb Space Telescope.

**Acknowledgements** We are grateful to L. Bergeron, S. Kassin, D. Magee, M. Stiavelli, R. Thompson and A. Zirm for their help in reducing the NICMOS data, to M. Franx for critical reading of this manuscript, to S. Malhotra for making the NICMOS parallels to her PEARS program public early, and to I. Labbe for assistance in interpreting the IRAC data.

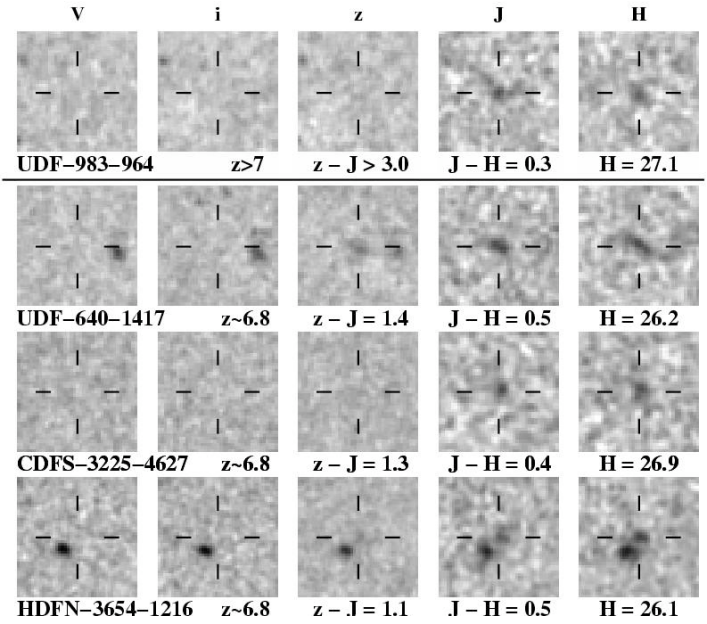

**Figure 1 | Optical and near-infrared images of four candidate galaxies at $z\approx7\text{-}8$.** The uppermost candidate is the only object satisfying our most conservative z-dropout criterion ($z_{850}-J_{110}>1.3$, $z_{850}-J_{110}>1.3+0.4(J_{110}-H_{160})$, $J_{110}-H_{160}<1.2$, undetected [$<2\sigma$] in the $V_{606}$ and $i_{775}$ bands). This object is undetected ($<2\sigma$) in the extremely deep optical B–V–i–z data available in the HUDF, has a $z_{850}-J_{110}$ colour redder than 3.0 mag ($1\sigma$), and a $J_{110}-H_{160}$ colour of 0.3 mag. It also shows a significant $5\sigma$ detection[31] in the InfraRed Array Channel (IRAC) 3.6-μm and 4.5-μm coverage (M. Dickinson *et al.*, manuscript in preparation) that we have from the Spitzer Space Telescope, with optical–infrared colours very similar to dropouts found at $z\approx6$ (refs 23–26). While this object lacks spectroscopic confirmation (and will probably continue to do so, because it is so faint), it seems extremely probable that it is at $z\approx7\text{-}8$. It had previously been reported as a z-dropout candidate in the HUDF[11]. The lower three candidates satisfy our less-conservative z-dropout criterion ($z_{850}-J_{110}>0.8$, $z_{850}-J_{110}>0.8+0.4(J_{110}-H_{160})$, $J_{110}-H_{160}<1.2$, undetected [$<2\sigma$] in the $V_{606}$ and $i_{775}$ bands). Though each of these sources is consistent with being at the low-redshift edge of the selection ($z\approx6.8$), the optical imaging available for two of these sources (CDFS−3225−4627 and HDFN−3654−1216) is not deep enough to completely rule out their being low-redshift sources (or T-dwarf stars in the case of CDFS−3225−4627). All z-dropout candidates considered here were detected at $>4\sigma$ and $>4.5\sigma$ in the J and H bands, respectively, and were found in NICMOS data with at least seven individual exposures. The V, i, z, J and H images shown here correspond to the ACS F606W, ACS F775W, ACS F850LP, NICMOS F110W, and NICMOS F160W filters, respectively, with central wavelengths of 591, 776, 944, 1,119 and 1,604 nm, respectively. Each cutout is 3.5″×3.5″ in size.

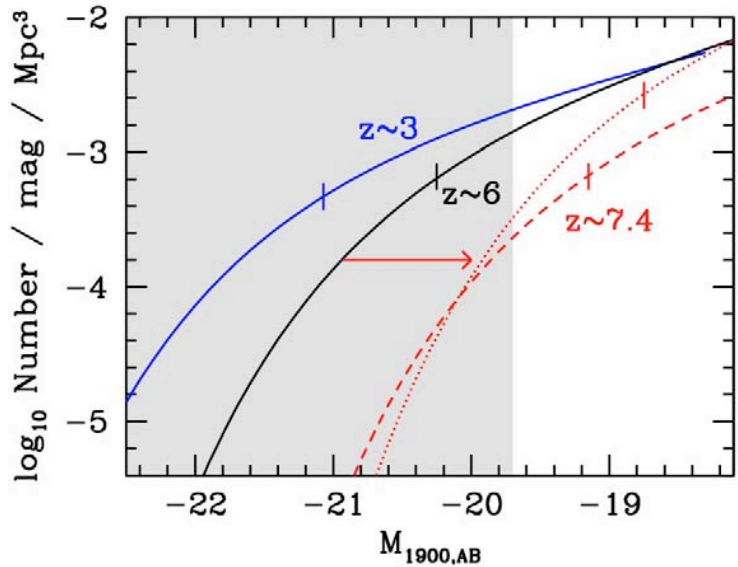

**Figure 2 | UV luminosity functions at redshift $z \approx 3$, 6 and 7.4.** The red lines show two possible luminosity functions at $z \approx 7-8$ that are consistent with the surface density of $z \approx 7-8$ galaxies (z-dropouts). These luminosity functions are compared with existing $z \approx 6$ and $z \approx 3$ luminosity functions[1,29]. The characteristic luminosity for each luminosity function is indicated on the figure with a short vertical line. The grey region shows the approximate flux range of the current z-dropout study. The luminosity functions here are plotted in terms of the number of galaxies per magnitude per cubic megaparsec and are a function of the absolute magnitude of galaxies ($M_{AB}$) at 1900 Å. The rest-frame UV luminosity functions brightens considerably from $z \approx 6$ to $z \approx 3$ (refs 1, 2), and it seems reasonable to imagine that this trend continues from even earlier times (that is, from $z \approx 7-8$ to $z \approx 6$). The first luminosity function (dashed line) was obtained by requiring the $z \approx 7-8$ luminosity function to have the same normalization $\phi^*$ and faint-end slope $\alpha$ as the luminosity function at $z \approx 6$ (solid black line)[1], but changing the characteristic luminosity $L^*$ to match the observed surface density of $z \approx 7-8$ galaxies. The second luminosity function (dotted line) was obtained by requiring the product of $\phi^*$ and $L^*$ to be a constant (to maintain an approximately constant luminosity density), fixing $\alpha$ to the $z \approx 6$ value ($\alpha = -1.73$), but changing the characteristic luminosity $L^*$ to match the observed surface density of $z \approx 7-8$ galaxies. The characteristic luminosity which best fits our search results is $1.1 \pm 0.4$ mag fainter than the $z \approx 6$ value for the first luminosity function and $1.5 \pm 0.4$ mag fainter for the second luminosity function (equivalent to $H_{160,AB}$-band magnitudes of $28.0 \pm 0.4$ mag and $28.4 \pm 0.4$ mag, respectively). This best-fit characteristic luminosity is 0.1 mag fainter and 0.1 mag brighter assuming faint-end slopes $\alpha$ of $-1.4$ and $-2.0$, respectively. The 'concordance' cosmology ($\Omega_m = 0.3$, $\Omega_\Lambda = 0.7$, $H_0 = 70$ km s$^{-1}$ Mpc$^3$) is assumed in this figure and in Fig. 3.

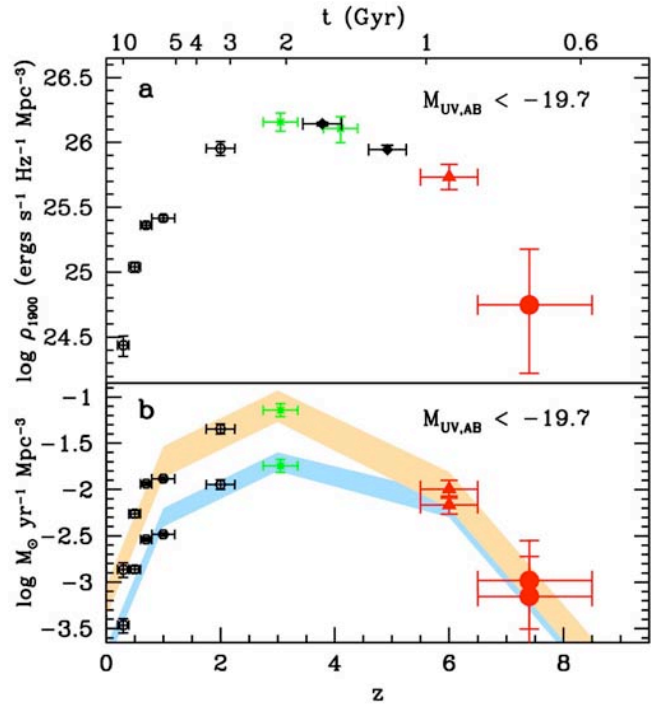

**Figure 3 | The luminosity density and star-formation rate density of the Universe over time (redshift). a**, The UV luminosity density at 1,900 Å from $z\approx0$ to $z\approx7$-8. Because of the flux limits of the current z-dropout searches, we plot the luminosity densities only down to a flux limit of $M_{AB} \approx -19.7$ mag (which is equivalent to 0.6 times the characteristic luminosity at $z=6$)[1]. The new determination is shown as a large filled circle at $z\approx7$-8 (with 1σ errors that include both poissonian and large-scale structure uncertainties) and compared with determinations at $z\approx0$–2 (open hexagons)[30], $z\approx3$ (green crosses)[29], $z\approx4$–5 (black diamonds)[22], and $z\approx6$ (red triangle)[1]. Vertical error bars on the latter determinations are all 1σ while the horizontal error bars show the approximate redshift range of applicability. The luminosity density shows an abrupt increase from $z\approx7$-8 to $z\approx6$, indicating that there has been a rapid build-up in luminous galaxies (those >0.6 times the characteristic luminosity at $z=6$, i.e., >0.6 $L^*_{z=6}$) over this range in redshift—which corresponds to a period of just 200 million years. **b**, The star-formation rate history of the Universe shown with and without a correction for dust extinction (upper and lower points, respectively). The errors on the points are the same as in the previous panel. This star-formation-rate history is plotted in terms of the number of solar-mass stars forming per year per unit comoving megaparsec and is derived here from the UV luminosity densities assuming a Salpeter initial mass function, which extends from $0.1 M_\odot$ to $100 M_\odot$, where $M_\odot$ is the mass of the Sun[32]. To account for the apparent evolution in the UV colours and thus the dust properties of high-redshift galaxies[1,26,33], we apply more minimal (~0.4 mag) dust corrections at $z>5$, but revert to more standard values (~1.4 mag) at $z\leq3$ (ref. 30).

# Supplementary Information: Rapid Evolution in the Most Luminous Galaxies During the First 900 Million Years

Rychard J. Bouwens, Garth D. Illingworth

Department of Astronomy, University of California Santa Cruz, 1156 High Street, Santa Cruz, CA 95064, USA

**Reductions of the Optical and Infrared (ACS/NICMOS) Data**

Our z-dropout search fields are shown in Supplementary Figure 1. In order to maximize the overall uniformity of the NICMOS data set, we re-reduced the majority of the data using our recently implemented NICMOS pipeline (R.J.B. *et al.*, in preparation). All exposures were subjected to the standard NICMOS processing procedures: flat fielding, bias and pedestal correction, cosmic ray rejection, bad pixel masking, treatment for SAA passage, and treatment for the "Mr. Staypuft" anomaly (which produces artificial sources 128 pixels from bright sources).[9] The individual exposures were then aligned and stacked, rejecting pixels more than 4σ outside the median. The only NICMOS data we did not re-reduce were those from the HDF-North, which were taken under very different operating conditions (prior to the refurbishment of the NICMOS camera in 2002, which changed its operating characteristics very significantly.)

For the ACS data, we made use of the public UDF reduction (S.D.W. Beckwith *et al.*, in preparation) and our own reduction of the UDF parallel and ACS GOODS data. The latter reduction included not only the original data from the ACS GOODS program,[13] but also the follow-up SNe search program (Riess *et al.*, in preparation) for a total of >9 orbits of $z_{850}$ band data at each position in the GOODS mosaic.[1] The ACS data were reduced with the ACS GTO pipeline "apsis" which handled the alignment, cosmic-ray rejection, and drizzling of the data onto a final output frame.[16]

**Selection Criteria for our z-dropout Samples**

The selection regions are shown in Supplementary Figure 2. For our primary, most conservative z-dropout selection, we required (1) $(z_{850}-J_{110})>1.3$, (2) $z_{850}-J_{110} > 1.3 + 0.4(J_{110}-H_{160})$, (3) $(J_{110}-H_{160})<1.2$, and (4) the candidates to be undetected (<2σ) in the $V_{606}i_{775}$ imaging. For the typical blue starburst galaxy at z>5 (with a UV continuum slope β ~ −2 where $f_\lambda \propto \lambda^\beta$), this selection probes objects over the range z~6.8-8.0. We chose a relatively stringent z–J cut to make our primary selection as clean as possible and largely exclude the region occupied by T-dwarfs (see Supplementary Figure 2).[34] Our $(J_{110}-H_{160})$ colour criterion was set to allow a good discrimination against redder galaxies at z~1-2, but still include bluer high-redshift objects out to redshifts as high as z~8. As an alternative to our primary selection, we also considered a selection with a weaker requirement on the $z_{850}$–$J_{110}$ colour to find candidate galaxies at slightly lower redshifts $z > \sim 6.5$ and to test the robustness of the results we derived from our primary selection. The criteria for this less-conservative selection were (1) $z_{850}$–$J_{110} > 0.8$, (2) $z_{850}$–$J_{110} > 0.8 + 0.4(J_{110}$–$H_{160})$, (3) $J_{110}$–$H_{160} < 1.2$, and (4) no detection ($< 2\sigma$) in the $V_{606}$ and $i_{775}$ bands. Note that in computing the z–J colour for both of our z-dropout selections, we set the z-band flux to its 1σ upper limit in case of a non-detection. Both criteria are illustrated on Supplementary Figure 2 with the photometric sample over-plotted.

**Conservative Selection**

Only two objects from our search fields satisfied our most conservative selection criteria. The first object has a ($z_{850}$-$J_{110}$) break greater than 3.0 mag (1σ), and a ($J_{110}$–$H_{160}$) colour of 0.3, indicative of a very blue UV continuum slope β ~ –2.5. The $H_{160,AB}$ magnitude is 27.1, corresponding to a UV luminosity of –20.0 AB mag. This object shows a clear (5σ) detection in the Spitzer Space Telescope IRAC (InfraRed Array Camera) 3.6μ and 4.5μ data,[29] with optical-IR colours consistent with i-dropouts found in the GOODS/UDF data.[23-26] Since its z–J break is >3.0 (1σ), it is too red to be a low-mass star, but far too blue (in its J–H colour) to be a low-redshift galaxy. Together these data indicate a very robust detection of a star forming galaxy at z > 7.3. This object had previously been reported as a z-dropout candidate in the UDF (UDF-983-964).[11] The second object in our selection (03:32:25.10, –27:46:35.6) has a z–J colour equal to 2.0, an $H_{160}$-band magnitude of 24.5, an J–H colour of 0.0, and is unresolved in the $z_{850}$ band. It is therefore very likely to be a T-dwarf (see Supplementary Figure 2).

**Less Conservative Selection**

There were four additional objects in our alternate selection, not present in our conservative selection. The first of these objects is a well-known high-redshift candidate from the UDF (UDF-640-1417);[3,11,36] it just missed the Bouwens et al. UDF z-dropout selection.[11] The second object from this selection (12:37:34.26, 62:18:31.4) again appears to be a T dwarf, being completely unresolved in the $z_{850}$ band. The nature of the two remaining objects from this selection (HDFN-3654-1216 and CDFS-3225-4627) was less easy to establish. Though they appear consistent with having redshifts just above z~6.5, we could not completely rule out their being lower redshift sources, or a T-dwarf in the case of CDFS-3225-4627. Of particular concern was HDFN-3654-1216, which together with its close neighbor, is somewhat bright in the public Spitzer IRAC data over the HDF North. The detection in the 3.6μ channel ($m_{3.6\mu,AB}$ ~ 24) suggests that it may be a low-redshift interloper.

**Comparison with our previous UDF z-dropout selection**

It is useful to compare our current z-dropout candidates from the UDF with those from our previous search.[11] In that work, we reported 5 possible z-dropout *candidates* over the UDF, but noted that there were concerns about possible contamination and instrumental effects. Our most conservative assumption at that time was that only a fraction of the detections were real. Using a new reduction of the NICMOS data over the UDF with a more conservative selection criterion, we find that only one of our original five candidates makes it into our present selection. Of the four candidates which miss our conservative selection, two (UDF-491-880 and UDF-818-886) are electronic ghosts of bright stars in the UDF ("the Mr. Staypuft effect").[9] Since the electronic ghosts of bright sources are now explicitly modeled and removed, such sources are not a concern for the current selection. The other two candidates (UDF-387-1125 and UDF-825-950), though evident in our reductions at lower significance, did not meet the more stringent selection requirements made here for either the conservative or the less conservative samples.

**Expected Number of z-dropouts Assuming No-Evolution From z~6**

We ran a series of Monte-Carlo simulations using our well-established cloning software, which enables us to generate realistic simulations of galaxies in the high-redshift universe.[18,19] Input catalogues for the z~6-10 universe were generated on the basis of the z~6 LF[1] and $V_{606}i_{775}z_{850}J_{110}H_{160}$ images generated by artificially redshifting similar luminosity B-dropouts (z~4) to the catalogue redshift, scaling the sizes as $(1+z)^{-1.1}$ (while keeping the luminosity fixed) to account for the expected size evolution and preserving the pixel-by-pixel morphologies.[1,35] The projected sample was assumed to have the same colour distribution as that determined at z~6 (i.e., a mean UV continuum slope β of −2.0 and 1σ scatter of 0.5).[1,26,33] The simulated images were then added to the ACS/NICMOS images which made up our search fields, and our selection procedure repeated. This latter step was particularly valuable given the rather heterogeneous nature of our search fields, with significant variations in the S/N of the optical and infrared data. The above simulations were repeated more than 20 times on each search field to minimize dependencies upon any particular realization.

These no-evolution simulations predicted that we should find ~10.1±3.2 galaxies over our search fields for our conservative z-dropout selection. 17.0±4.1 galaxies were predicted for the less-conservative selection criteria, again if there was no evolution in the UV LF from z~7 to z~6. Incorporating the uncertainties in the z~6 LF which result from large-scale structure and small number statistics (together these are estimated to make the overall normalization of the z~6 LF uncertain at the ~16% level),[1] the no-evolution predictions for our two selections become ~10.1±3.5 and 17.0±4.9, respectively. To determine the dependence of these predictions on the assumed sizes, we repeated the above simulations assuming the sizes were 30% larger and 30% smaller. For these assumptions, our derived numbers were 36% smaller and 24% larger, respectively. Our simulations suggested that our z-dropout selections should identify star-forming galaxies over the redshift range z~6.5 to z~8.5, with a mean value of z~7.4. Most of the $z_{850}$-dropouts recovered in the simulations had $H_{160,AB}$ magnitudes ranging from 26.0 to 27.5, suggesting that the current search was most effective at probing absolute magnitudes between –19.7 and –21.2 AB mag (or 0.6-2.5 L*(z=6)).

**Field-to-Field Variations**

Current theories for the growth of structure allow us to estimate the impact of the clustering provided we have a good estimate of the bias, which provides a link between the observed galaxies and the underlying fluctuations in the mass. It has become conventional to calculate the bias for a population by considering the number density of a population, converting this to a mass scale and then calculating the equivalent bias.[20] For the current search which probes to volume densities of ~$10^{-4}$ Mpc$^{-3}$, this corresponds to a bias of 7.[20,21] Assuming a pencil-beam geometry and selection window of width $\Delta z = 1$ at z~7-8, we estimate the RMS variance in the number density of $z_{850}$-dropouts in each of our search fields. This variance is 49% for single 0.8 arcmin$^2$ Near Infrared Camera 3 (NIC3) pointings and 41% for larger ~5 arcmin$^2$ search fields like the UDF or HDF North NICMOS mosaics. Since the entire CDF-S or HDF-N GOODS field (150 arcmin$^2$) has an estimated RMS variance of ~24%, approximately 30% of the variance in our individual NICMOS fields is coherent across the larger field. Field-to-field variance is modeled using a log-normal distribution.

**Star Formation Rate Density**

The current measure of the rest-frame UV LF at z~7.4 can be used to estimate of the rest-frame UV luminosity density and star formation rate densities to the approximate magnitude limit of the probe ($M_{UV,AB}$ = –19.7, or 0.6 L*(z=6)). Using the luminosity density at z~6 down to this limit and multiplying by the evolutionary factor $0.10^{+0.19}_{-0.07}$ × derived from our most conservative selection, we find a rest-frame (1900 Å) UV luminosity density of $5.4^{+9.2}_{-3.8}$ x $10^{24}$ erg/cm$^2$/s/Mpc$^3$. For a Salpeter initial mass function with an assumed mass range 0.1 – 100 $M_{sol}$, this luminosity density translates into an unextincted star formation rate (SFR) density of $6.7^{+11.2}_{-4.8}$ x $10^{-4}$ $M_o$ Mpc$^{-3}$ yr$^{-1}$.[32] We plot the current determination in Figure 3 and compare it against previous measures of the SFR density over the range 0≤z≤7.4.[1,22,30,31] There is a clear, almost precipitous rise in the UV luminosity density from z~7.4 to z~6. However, it is important to remember that this is just the evolution of the SFR density evaluated down to a relatively bright flux limit (–19.7 AB mag or 0.6 L*(z=6)). At z~6, the LF is known to be steep with substantial contributions to the UV flux density from faint galaxies.[1,2,3] Since galaxies in general are likely to be less massive (and hence less luminous) at z~7.4 than at z~6, it seems probable that a substantial fraction of the total star formation at z~7.4 is also below the 0.6L* limit, and that the evolution in the total SFR density is much shallower than shown in Figure 3.

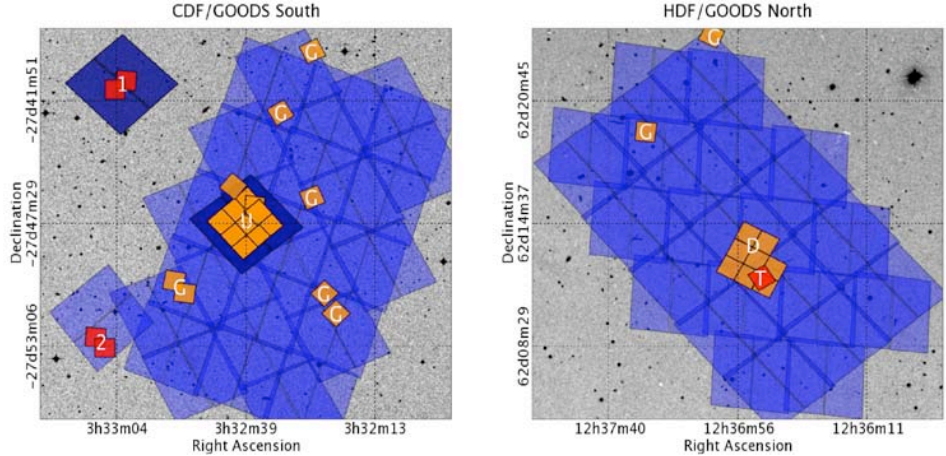

Supplementary Figure 1. Search fields used to find z~7-8 z-dropouts. The left and right panels (18 arcmin on a side) show the CDF-S and HDF-N GOODS fields, respectively. Our NICMOS search areas are labeled ("U"="UDF", "1"="UDF NICPAR1", "2"="UDF NICPAR2", "D"="Dickinson", "T"="Thompson", and "G"="GOODS") to allow easy identification with the fields listed in Supplementary Table 1. These search areas are colour-coded according to their depth (darker is deeper) to give our readers a sense for the available data. The orange and red regions correspond to NICMOS pointings with 5σ depths of >27 and >28, respectively, in $J_{110}$ and $H_{160}$. The dark blue regions show the ultra deep ACS data available over the UDF and UDF NICMOS parallel field #1. The light blue regions show the area covered by the deep *Viz* ACS data over the GOODS fields and UDF NICMOS parallel field #2. Our strongest z~7-8 candidate (conservative selection) is found near the "U" in the UDF. Coordinate labels (J2000 equinox) are included along the vertical and horizontal axes to indicate the position of these fields on the sky.

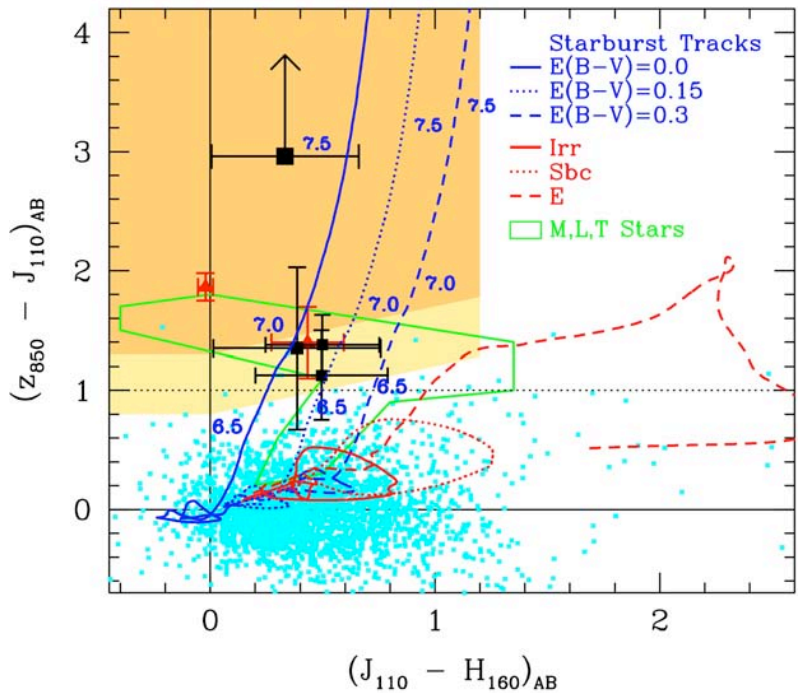

Supplementary Figure 2. $(z_{850} - J_{110})_{AB}$ / $(J_{110} - H_{160})_{AB}$ colour-colour diagram used for selecting our high-redshift z-dropout sample. The selection window for our primary z-dropout sample is shown with the darker orange shading and only includes one plausible z>7 candidate (shown as large black square with an arrow showing the 1σ lower limit on its $z_{850}$–$J_{110}$ colour). The selection window for our other z-dropout sources is shown with the light orange shading and includes three plausible z>6.5 candidates (shown as the smaller black squares with 1σ errors). Lower-redshift sources detected (>2σ) in the HST ACS $V_{606}$ or $i_{775}$ band images are plotted in cyan. The red triangles mark the position of two probable T-dwarfs. The objects shown here are from the UDF, HDF-North, the UDF NICMOS Parallels, and other NICMOS parallels over the GOODS field (Supplementary Table 1 and Supplementary Figure 1). Colour-colour tracks for $10^8$ yr constant star formation models with different amounts of dust extinction (E(B-V)) are shown with the blue lines to indicate the likely position of z>6.5 candidates. Similar tracks are also included for several low-redshift galaxy spectral templates (red lines). Lower mass stars are expected to lie within the region demarcated by the green line.[34]

**Supplementary Table 1. z-dropout search fields**

| Field | Area (arcmin$^2$) | $z_{850}$ Depth | $J_{110}$ Depth | $H_{160}$ Depth |
|---|---|---|---|---|
| UDF[10] | 6.5 | 29.0 | 27.7 | 27.5 |
| UDF NICPAR 1 | 1.3 | 28.6 | 28.7 | 28.5 |
| UDF NICPAR 2 | 1.3 | 27.5 | 28.7 | 28.5 |
| GOODS | 4.9 | 27.5 | 27.2 | 27.0 |
| HDF-North Thompson[14] | 0.8 | 27.8 | 28.0 | 28.1 |
| HDF-North Dickinson[15] | 4.0 | 27.8 | 27.0 | 27.0 |

The depths quoted are the typical 5σ depths in 0.6"-diameter aperture in the $z_{850}$, $J_{110}$, and $H_{160}$ bands over those areas. Here the GOODS search fields include all areas over the ACS GOODS fields (excluding the UDF or HDF-North) with deep NICMOS observations. Supplementary Figure 1 shows the position and general layout of the above fields. The NICMOS data for the "UDF NICPAR 1" and "UDF NICPAR 2" fields were taken in parallel to the ACS HUDF observations (S.D.W. Beckwith *et al.*, in preparation), while the ACS data come from follow-up observations (M. Stiavelli *et al.*, in preparation). The FWHM of the PSF for the NIC3 J and H data is ~0.35", except for the case of the HDF-North data where it was ~0.25". Note that for the HDF-North WFPC2 field where coverage is available from both the Thompson and Dickinson programs (Supplementary Figure 1), we used the deeper HDF-N Thompson data for z-dropout searches over that 0.8 arcmin$^2$ field and the Dickinson data elsewhere. We have quoted a correspondingly smaller area for the HDF-N Dickinson mosaic to account for the overlap. Also note that for the few regions where there is a mismatch between the depths of our optical and infrared data (most notably in the case of the UDF NICPAR2 field), our z-dropout search is limited by the depth of the shallower ACS data.

**Supplementary Table 2. z-dropout (z>6.5) candidates**

| Object ID | Right Ascension | Declination | $H_{160,AB}$ | $z_{850}-J_{110}$ | $J_{110}-H_{160}$ | Ref |
|---|---|---|---|---|---|---|
| UDF-983-964** | 03:32:38.79 | −27:47:07.1 | 27.0 ± 0.2 | >3.0 | 0.3 ± 0.3 | [11,36] |
| UDF-640-1417 | 03:32:42.56 | −27:46:56.6 | 26.2 ± 0.1 | 1.4 ± 0.2 | 0.5 ± 0.2 | [3,11,36] |
| CDFS-3225-4627 | 03:32:25.22 | −27:46:26.6 | 26.9 ± 0.2 | 1.3 ± 0.6 | 0.4 ± 0.4 | ---- |
| HDFN-3654-1216 | 12:36:54.12 | 62:12:16.2 | 26.1 ± 0.1 | 1.1 ± 0.4 | 0.5 ± 0.3 | ---- |

All coordinates are for the J2000 equinox. Total magnitudes here are quoted in the $H_{160}$ band and were measured in large scalable apertures (with typical diameters of 1.1").[37] Colours were measured within a small scalable aperture (with typical diameters of 0.5"). The uppermost object in this table (UDF-983-964**) is the only source in our most conservative z-dropout selection ($z_{850}-J_{110}$ > 1.3, $z_{850}-J_{110}$ > 1.3 + 0.4($J_{110}-H_{160}$), $J_{110}-H_{160}$<1.2, undetected [<2σ] in $V_{606}$ and $i_{775}$ bands). All four sources in this table are in our less conservative z-dropout selection ($z_{850}-J_{110}$ > 0.8, $z_{850}-J_{110}$ > 0.8 + 0.4($J_{110}-H_{160}$), $J_{110}-H_{160}$<1.2, undetected [<2σ] in $V_{606}$ and $i_{775}$ bands).